\documentstyle[12pt]{article}
\begin{document}
\title{\bf On  Masses of Equilibrium Configurations}
\author{L.V.Verozub}
\maketitle
\centerline{\em Department of Physics and Astronomy, Kharkov State University}
\centerline{\em Kharkov 310077 Ukraine}
\begin{abstract}
{\em Proceeding from the new gravitation equations (Phys.Lett.A, v.156, p.404
(1991) ) we argue that the  theory in principle allows equilibrium 
stable configurations of a degenerate electron or neutron gas with very large 
masses.}
\end{abstract}

Proceeding from the newtonian gravity law and Einstein's equations
it is considered  that masses of equilibrium configurations cannot go
over several Sun masses. In paper \cite{ver} new vacuum gravitational 
equations in
flat space-time was proposed, which have no  physical singularity for the
spherically symmetric field. If the distance $r$ from an attractive mass $M$
is much larger than  the Shwarzshild radius $\alpha$, then their 
physical consequences  coinside with  the ones in  Einstein's theory.
However, they are quite different at $r$ of the order $\alpha$ or less
than that.
  There is no events horizon at $r=\alpha$ . The gravitational force
affecting the a test particle of the mass $m$ in rest is given by

\begin{equation}
 F=-\frac{GmM}{r^2} \left(1-\alpha/f) \right.,  
\end{equation}

where $G$ is the gravitational constant, $\alpha=2GM/c^{2}$, $c$ is the speed
of light, $f=(\alpha^3+r^3)^{1/3}$ . Fig. 1 shows the plot of the function
$F_{1} = - (1/2\  \overline r^{2}) (1 - \alpha /f)$  against the distance 
$\overline r=r/ \alpha$. 

\setlength{\unitlength}{0.240900pt}
\ifx\plotpoint\undefined\newsavebox{\plotpoint}\fi
\sbox{\plotpoint}{\rule[-0.200pt]{0.400pt}{0.400pt}}%
\special{em:linewidth 0.4pt}%
\begin{picture}(1500,900)(0,0)
\tenrm
\put(264,158){\special{em:moveto}}
\put(264,787){\special{em:lineto}}
\put(264,158){\special{em:moveto}}
\put(284,158){\special{em:lineto}}
\put(1436,158){\special{em:moveto}}
\put(1416,158){\special{em:lineto}}
\put(242,158){\makebox(0,0)[r]{-0.22}}
\put(264,215){\special{em:moveto}}
\put(284,215){\special{em:lineto}}
\put(1436,215){\special{em:moveto}}
\put(1416,215){\special{em:lineto}}
\put(242,215){\makebox(0,0)[r]{-0.2}}
\put(264,272){\special{em:moveto}}
\put(284,272){\special{em:lineto}}
\put(1436,272){\special{em:moveto}}
\put(1416,272){\special{em:lineto}}
\put(242,272){\makebox(0,0)[r]{-0.18}}
\put(264,330){\special{em:moveto}}
\put(284,330){\special{em:lineto}}
\put(1436,330){\special{em:moveto}}
\put(1416,330){\special{em:lineto}}
\put(242,330){\makebox(0,0)[r]{-0.16}}
\put(264,387){\special{em:moveto}}
\put(284,387){\special{em:lineto}}
\put(1436,387){\special{em:moveto}}
\put(1416,387){\special{em:lineto}}
\put(242,387){\makebox(0,0)[r]{-0.14}}
\put(264,444){\special{em:moveto}}
\put(284,444){\special{em:lineto}}
\put(1436,444){\special{em:moveto}}
\put(1416,444){\special{em:lineto}}
\put(242,444){\makebox(0,0)[r]{-0.12}}
\put(264,501){\special{em:moveto}}
\put(284,501){\special{em:lineto}}
\put(1436,501){\special{em:moveto}}
\put(1416,501){\special{em:lineto}}
\put(242,501){\makebox(0,0)[r]{-0.1}}
\put(264,558){\special{em:moveto}}
\put(284,558){\special{em:lineto}}
\put(1436,558){\special{em:moveto}}
\put(1416,558){\special{em:lineto}}
\put(242,558){\makebox(0,0)[r]{-0.08}}
\put(264,615){\special{em:moveto}}
\put(284,615){\special{em:lineto}}
\put(1436,615){\special{em:moveto}}
\put(1416,615){\special{em:lineto}}
\put(242,615){\makebox(0,0)[r]{-0.06}}
\put(264,673){\special{em:moveto}}
\put(284,673){\special{em:lineto}}
\put(1436,673){\special{em:moveto}}
\put(1416,673){\special{em:lineto}}
\put(242,673){\makebox(0,0)[r]{-0.04}}
\put(264,730){\special{em:moveto}}
\put(284,730){\special{em:lineto}}
\put(1436,730){\special{em:moveto}}
\put(1416,730){\special{em:lineto}}
\put(242,730){\makebox(0,0)[r]{-0.02}}
\put(264,787){\special{em:moveto}}
\put(284,787){\special{em:lineto}}
\put(1436,787){\special{em:moveto}}
\put(1416,787){\special{em:lineto}}
\put(242,787){\makebox(0,0)[r]{0}}
\put(264,158){\special{em:moveto}}
\put(264,178){\special{em:lineto}}
\put(264,787){\special{em:moveto}}
\put(264,767){\special{em:lineto}}
\put(264,113){\makebox(0,0){0}}
\put(498,158){\special{em:moveto}}
\put(498,178){\special{em:lineto}}
\put(498,787){\special{em:moveto}}
\put(498,767){\special{em:lineto}}
\put(498,113){\makebox(0,0){2}}
\put(733,158){\special{em:moveto}}
\put(733,178){\special{em:lineto}}
\put(733,787){\special{em:moveto}}
\put(733,767){\special{em:lineto}}
\put(733,113){\makebox(0,0){4}}
\put(967,158){\special{em:moveto}}
\put(967,178){\special{em:lineto}}
\put(967,787){\special{em:moveto}}
\put(967,767){\special{em:lineto}}
\put(967,113){\makebox(0,0){6}}
\put(1202,158){\special{em:moveto}}
\put(1202,178){\special{em:lineto}}
\put(1202,787){\special{em:moveto}}
\put(1202,767){\special{em:lineto}}
\put(1202,113){\makebox(0,0){8}}
\put(1436,158){\special{em:moveto}}
\put(1436,178){\special{em:lineto}}
\put(1436,787){\special{em:moveto}}
\put(1436,767){\special{em:lineto}}
\put(1436,113){\makebox(0,0){10}}
\put(264,158){\special{em:moveto}}
\put(1436,158){\special{em:lineto}}
\put(1436,787){\special{em:lineto}}
\put(264,787){\special{em:lineto}}
\put(264,158){\special{em:lineto}}
\put(850,68){\makebox(0,0){$\overline r$}}
\put(381,730){\makebox(0,0)[r]{$F_{1}$}}
\put(403,730){\special{em:moveto}}
\put(469,730){\special{em:lineto}}
\put(276,692){\special{em:moveto}}
\put(287,597){\special{em:lineto}}
\put(299,506){\special{em:lineto}}
\put(311,421){\special{em:lineto}}
\put(323,347){\special{em:lineto}}
\put(334,286){\special{em:lineto}}
\put(346,241){\special{em:lineto}}
\put(358,212){\special{em:lineto}}
\put(369,198){\special{em:lineto}}
\put(381,197){\special{em:lineto}}
\put(393,206){\special{em:lineto}}
\put(405,222){\special{em:lineto}}
\put(416,244){\special{em:lineto}}
\put(428,268){\special{em:lineto}}
\put(440,293){\special{em:lineto}}
\put(452,319){\special{em:lineto}}
\put(463,345){\special{em:lineto}}
\put(475,370){\special{em:lineto}}
\put(487,393){\special{em:lineto}}
\put(498,416){\special{em:lineto}}
\put(510,437){\special{em:lineto}}
\put(522,457){\special{em:lineto}}
\put(534,475){\special{em:lineto}}
\put(545,493){\special{em:lineto}}
\put(557,509){\special{em:lineto}}
\put(569,524){\special{em:lineto}}
\put(580,538){\special{em:lineto}}
\put(592,551){\special{em:lineto}}
\put(604,563){\special{em:lineto}}
\put(616,574){\special{em:lineto}}
\put(627,584){\special{em:lineto}}
\put(639,594){\special{em:lineto}}
\put(651,603){\special{em:lineto}}
\put(662,612){\special{em:lineto}}
\put(674,620){\special{em:lineto}}
\put(686,627){\special{em:lineto}}
\put(698,634){\special{em:lineto}}
\put(709,641){\special{em:lineto}}
\put(721,647){\special{em:lineto}}
\put(733,653){\special{em:lineto}}
\put(745,658){\special{em:lineto}}
\put(756,663){\special{em:lineto}}
\put(768,668){\special{em:lineto}}
\put(780,673){\special{em:lineto}}
\put(791,677){\special{em:lineto}}
\put(803,681){\special{em:lineto}}
\put(815,685){\special{em:lineto}}
\put(827,689){\special{em:lineto}}
\put(838,692){\special{em:lineto}}
\put(850,695){\special{em:lineto}}
\put(862,699){\special{em:lineto}}
\put(873,702){\special{em:lineto}}
\put(885,704){\special{em:lineto}}
\put(897,707){\special{em:lineto}}
\put(909,710){\special{em:lineto}}
\put(920,712){\special{em:lineto}}
\put(932,714){\special{em:lineto}}
\put(944,717){\special{em:lineto}}
\put(955,719){\special{em:lineto}}
\put(967,721){\special{em:lineto}}
\put(979,723){\special{em:lineto}}
\put(991,725){\special{em:lineto}}
\put(1002,726){\special{em:lineto}}
\put(1014,728){\special{em:lineto}}
\put(1026,730){\special{em:lineto}}
\put(1038,731){\special{em:lineto}}
\put(1049,733){\special{em:lineto}}
\put(1061,734){\special{em:lineto}}
\put(1073,736){\special{em:lineto}}
\put(1084,737){\special{em:lineto}}
\put(1096,738){\special{em:lineto}}
\put(1108,740){\special{em:lineto}}
\put(1120,741){\special{em:lineto}}
\put(1131,742){\special{em:lineto}}
\put(1143,743){\special{em:lineto}}
\put(1155,744){\special{em:lineto}}
\put(1166,745){\special{em:lineto}}
\put(1178,746){\special{em:lineto}}
\put(1190,747){\special{em:lineto}}
\put(1202,748){\special{em:lineto}}
\put(1213,749){\special{em:lineto}}
\put(1225,750){\special{em:lineto}}
\put(1237,750){\special{em:lineto}}
\put(1248,751){\special{em:lineto}}
\put(1260,752){\special{em:lineto}}
\put(1272,753){\special{em:lineto}}
\put(1284,754){\special{em:lineto}}
\put(1295,754){\special{em:lineto}}
\put(1307,755){\special{em:lineto}}
\put(1319,756){\special{em:lineto}}
\put(1331,756){\special{em:lineto}}
\put(1342,757){\special{em:lineto}}
\put(1354,757){\special{em:lineto}}
\put(1366,758){\special{em:lineto}}
\put(1377,759){\special{em:lineto}}
\put(1389,759){\special{em:lineto}}
\put(1401,760){\special{em:lineto}}
\put(1413,760){\special{em:lineto}}
\put(1424,761){\special{em:lineto}}
\put(1436,761){\special{em:lineto}}
\end{picture}

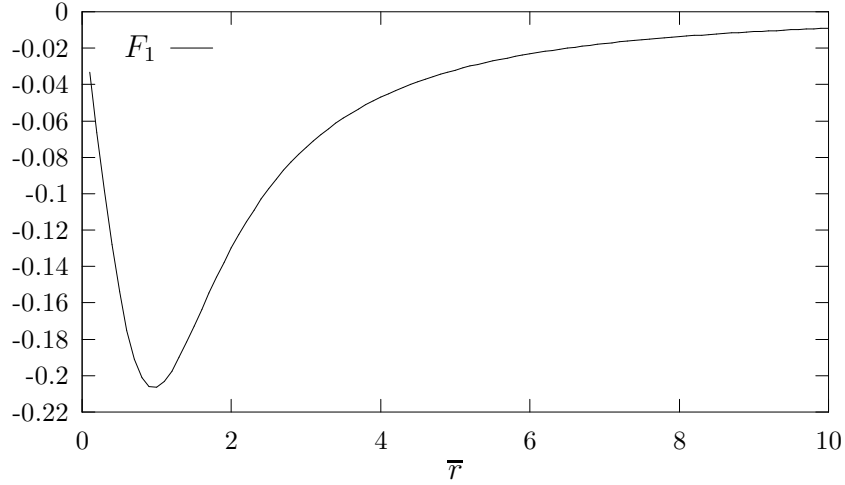
Fig.1  The plot of the function $F_{1}$ against the $\overline r = r/ \alpha$
\vskip 1cm

It follows from Fig.1 that the $|F|$ reaches its maximum at $r$ of the order
of $\alpha$ and tends to zero at $ r\rightarrow 0 $.
It would therefore be interesting to khow what the limiting masses of
the equilibrium configurations the gravitational force $F(r)$\ (1) can admit.
To answer this question we start from the equation
\begin{equation}
  \frac{dp}{dr} = -\frac{G \rho M}{r^2} \left(1-\alpha/f) \right.
\end{equation}
In this equation $\rho$ is the pressure , $M=M(r)$ is the matter mass inside
of a sphere of the radious $r$, $\rho=\rho(r)$ is the matter density at the 
distance $r$ from the center, $\alpha$ and $f$ is the function of $M(r)$.

Suppose the equation of state is $p=K\rho^\Gamma$, where  $K$ and $\Gamma$
are constants. For numerical estimates we shall use  
their values \cite{teuk} :

For a degenerated electron gas:

$\Gamma=5/3$ \   $K=1\cdot 10^{13}$ SGS units at $\rho\ll\rho_{0}$, where 
$\rho_{0}=10^{6}\   gm/cm^{3} $,

$\Gamma=4/3$ \   $K=1\cdot 10^{15}$ SGS units at  $\rho\gg\rho_{0}$.

For degenerated neutron gas:

$\Gamma=5/3$ \ $K=5\cdot 10^{9}$ SGS units at $\rho\ll\rho_{0}$, where
$\rho_{0}=5\cdot 10^{15}  gm/cm^{3} $ ,

$\Gamma=4/3$ \ $K=1\cdot 10^{15}$ SGS units at $\rho\gg\rho_{0}$.

for rough estimates   we replace $dp/dr$ by $-p/r$, where $p$ is
the average matter pressure and $R$ is its radius. Under the circumstances
we obtain from eq.(2) 
\begin{equation}
  \frac{p}{\rho c^{2}} = \frac{\alpha}{2R} \left (1-\alpha/f)\right. .
\end{equation}

If $ R \gg \alpha $ , then the term $\alpha /f$ is negligible. Setting 
      $M(R)\approx \rho R^{3}$ we find the mass of equilibrium states as a 
function of $\rho$ :
\begin{equation}
    M=(K/G)^{3/2} \rho ^{ (\Gamma - 4/3)(3/2)}  .     
\end{equation}

It follows from eq.(4) that there is the maximal mass  \cite{chan}
$M=(K/G)^{3/2}$ at $\rho \gg \rho _{0} $ .

However, according to eq.(3) , there are also equilibrium configurations  at
$R < \alpha$ . In particular, at $R \ll \alpha $ we find from eq.(3) that
the masses of the equilibrium configurations are given by
\begin{equation}
 M= c^{9/2} 10 ^{-1} K^{-3/4} G^{-3/2} \rho ^{-(\Gamma -1/3)(3/4)} . 
\end{equation}
These are the configurations with very large masses. For example, the 
following equilibrium configurations can be found:

the nonrelativistic electrons: $\rho =10^{5} gm/cm^{3}$, 
$M=1.3\cdot 10^{42} gm$, $R=2,3\cdot 10^{12}  cm$, 

the relativistic electrons: $\rho = 10^{7} gm/cm^{3}$, 
$M=2.3\cdot10^{40} gm$, $R=1.3\cdot 10^{11} cm$,

the nonrelativistic neutrons: $\rho =10^{14} gm/cm^{3}$
$M=3.9\cdot10^{35} gm$, $R=1.6\cdot 10^{7} cm$.

The reason of the two types of configurations existence can be seen from Fig. 
2, where for $\rho =10^{15}\ gm/cm^{3}$  the plots of right-hand and left-hand
sides of Eq.(3) against  the mass $ M$ are given. 

\setlength{\unitlength}{0.240900pt}
\ifx\plotpoint\undefined\newsavebox{\plotpoint}\fi
\sbox{\plotpoint}{\rule[-0.200pt]{0.400pt}{0.400pt}}%
\special{em:linewidth 0.4pt}%
\begin{picture}(1500,900)(0,0)
\tenrm
\put(264,158){\special{em:moveto}}
\put(264,787){\special{em:lineto}}
\put(264,158){\special{em:moveto}}
\put(284,158){\special{em:lineto}}
\put(1436,158){\special{em:moveto}}
\put(1416,158){\special{em:lineto}}
\put(242,158){\makebox(0,0)[r]{-0.22}}
\put(264,215){\special{em:moveto}}
\put(284,215){\special{em:lineto}}
\put(1436,215){\special{em:moveto}}
\put(1416,215){\special{em:lineto}}
\put(242,215){\makebox(0,0)[r]{-0.2}}
\put(264,272){\special{em:moveto}}
\put(284,272){\special{em:lineto}}
\put(1436,272){\special{em:moveto}}
\put(1416,272){\special{em:lineto}}
\put(242,272){\makebox(0,0)[r]{-0.18}}
\put(264,330){\special{em:moveto}}
\put(284,330){\special{em:lineto}}
\put(1436,330){\special{em:moveto}}
\put(1416,330){\special{em:lineto}}
\put(242,330){\makebox(0,0)[r]{-0.16}}
\put(264,387){\special{em:moveto}}
\put(284,387){\special{em:lineto}}
\put(1436,387){\special{em:moveto}}
\put(1416,387){\special{em:lineto}}
\put(242,387){\makebox(0,0)[r]{-0.14}}
\put(264,444){\special{em:moveto}}
\put(284,444){\special{em:lineto}}
\put(1436,444){\special{em:moveto}}
\put(1416,444){\special{em:lineto}}
\put(242,444){\makebox(0,0)[r]{-0.12}}
\put(264,501){\special{em:moveto}}
\put(284,501){\special{em:lineto}}
\put(1436,501){\special{em:moveto}}
\put(1416,501){\special{em:lineto}}
\put(242,501){\makebox(0,0)[r]{-0.1}}
\put(264,558){\special{em:moveto}}
\put(284,558){\special{em:lineto}}
\put(1436,558){\special{em:moveto}}
\put(1416,558){\special{em:lineto}}
\put(242,558){\makebox(0,0)[r]{-0.08}}
\put(264,615){\special{em:moveto}}
\put(284,615){\special{em:lineto}}
\put(1436,615){\special{em:moveto}}
\put(1416,615){\special{em:lineto}}
\put(242,615){\makebox(0,0)[r]{-0.06}}
\put(264,673){\special{em:moveto}}
\put(284,673){\special{em:lineto}}
\put(1436,673){\special{em:moveto}}
\put(1416,673){\special{em:lineto}}
\put(242,673){\makebox(0,0)[r]{-0.04}}
\put(264,730){\special{em:moveto}}
\put(284,730){\special{em:lineto}}
\put(1436,730){\special{em:moveto}}
\put(1416,730){\special{em:lineto}}
\put(242,730){\makebox(0,0)[r]{-0.02}}
\put(264,787){\special{em:moveto}}
\put(284,787){\special{em:lineto}}
\put(1436,787){\special{em:moveto}}
\put(1416,787){\special{em:lineto}}
\put(242,787){\makebox(0,0)[r]{0}}
\put(264,158){\special{em:moveto}}
\put(264,178){\special{em:lineto}}
\put(264,787){\special{em:moveto}}
\put(264,767){\special{em:lineto}}
\put(264,113){\makebox(0,0){0}}
\put(498,158){\special{em:moveto}}
\put(498,178){\special{em:lineto}}
\put(498,787){\special{em:moveto}}
\put(498,767){\special{em:lineto}}
\put(498,113){\makebox(0,0){2}}
\put(733,158){\special{em:moveto}}
\put(733,178){\special{em:lineto}}
\put(733,787){\special{em:moveto}}
\put(733,767){\special{em:lineto}}
\put(733,113){\makebox(0,0){4}}
\put(967,158){\special{em:moveto}}
\put(967,178){\special{em:lineto}}
\put(967,787){\special{em:moveto}}
\put(967,767){\special{em:lineto}}
\put(967,113){\makebox(0,0){6}}
\put(1202,158){\special{em:moveto}}
\put(1202,178){\special{em:lineto}}
\put(1202,787){\special{em:moveto}}
\put(1202,767){\special{em:lineto}}
\put(1202,113){\makebox(0,0){8}}
\put(1436,158){\special{em:moveto}}
\put(1436,178){\special{em:lineto}}
\put(1436,787){\special{em:moveto}}
\put(1436,767){\special{em:lineto}}
\put(1436,113){\makebox(0,0){10}}
\put(264,158){\special{em:moveto}}
\put(1436,158){\special{em:lineto}}
\put(1436,787){\special{em:lineto}}
\put(264,787){\special{em:lineto}}
\put(264,158){\special{em:lineto}}
\put(850,68){\makebox(0,0){$\overline r$}}
\put(381,730){\makebox(0,0)[r]{$F_{1}$}}
\put(403,730){\special{em:moveto}}
\put(469,730){\special{em:lineto}}
\put(276,692){\special{em:moveto}}
\put(287,597){\special{em:lineto}}
\put(299,506){\special{em:lineto}}
\put(311,421){\special{em:lineto}}
\put(323,347){\special{em:lineto}}
\put(334,286){\special{em:lineto}}
\put(346,241){\special{em:lineto}}
\put(358,212){\special{em:lineto}}
\put(369,198){\special{em:lineto}}
\put(381,197){\special{em:lineto}}
\put(393,206){\special{em:lineto}}
\put(405,222){\special{em:lineto}}
\put(416,244){\special{em:lineto}}
\put(428,268){\special{em:lineto}}
\put(440,293){\special{em:lineto}}
\put(452,319){\special{em:lineto}}
\put(463,345){\special{em:lineto}}
\put(475,370){\special{em:lineto}}
\put(487,393){\special{em:lineto}}
\put(498,416){\special{em:lineto}}
\put(510,437){\special{em:lineto}}
\put(522,457){\special{em:lineto}}
\put(534,475){\special{em:lineto}}
\put(545,493){\special{em:lineto}}
\put(557,509){\special{em:lineto}}
\put(569,524){\special{em:lineto}}
\put(580,538){\special{em:lineto}}
\put(592,551){\special{em:lineto}}
\put(604,563){\special{em:lineto}}
\put(616,574){\special{em:lineto}}
\put(627,584){\special{em:lineto}}
\put(639,594){\special{em:lineto}}
\put(651,603){\special{em:lineto}}
\put(662,612){\special{em:lineto}}
\put(674,620){\special{em:lineto}}
\put(686,627){\special{em:lineto}}
\put(698,634){\special{em:lineto}}
\put(709,641){\special{em:lineto}}
\put(721,647){\special{em:lineto}}
\put(733,653){\special{em:lineto}}
\put(745,658){\special{em:lineto}}
\put(756,663){\special{em:lineto}}
\put(768,668){\special{em:lineto}}
\put(780,673){\special{em:lineto}}
\put(791,677){\special{em:lineto}}
\put(803,681){\special{em:lineto}}
\put(815,685){\special{em:lineto}}
\put(827,689){\special{em:lineto}}
\put(838,692){\special{em:lineto}}
\put(850,695){\special{em:lineto}}
\put(862,699){\special{em:lineto}}
\put(873,702){\special{em:lineto}}
\put(885,704){\special{em:lineto}}
\put(897,707){\special{em:lineto}}
\put(909,710){\special{em:lineto}}
\put(920,712){\special{em:lineto}}
\put(932,714){\special{em:lineto}}
\put(944,717){\special{em:lineto}}
\put(955,719){\special{em:lineto}}
\put(967,721){\special{em:lineto}}
\put(979,723){\special{em:lineto}}
\put(991,725){\special{em:lineto}}
\put(1002,726){\special{em:lineto}}
\put(1014,728){\special{em:lineto}}
\put(1026,730){\special{em:lineto}}
\put(1038,731){\special{em:lineto}}
\put(1049,733){\special{em:lineto}}
\put(1061,734){\special{em:lineto}}
\put(1073,736){\special{em:lineto}}
\put(1084,737){\special{em:lineto}}
\put(1096,738){\special{em:lineto}}
\put(1108,740){\special{em:lineto}}
\put(1120,741){\special{em:lineto}}
\put(1131,742){\special{em:lineto}}
\put(1143,743){\special{em:lineto}}
\put(1155,744){\special{em:lineto}}
\put(1166,745){\special{em:lineto}}
\put(1178,746){\special{em:lineto}}
\put(1190,747){\special{em:lineto}}
\put(1202,748){\special{em:lineto}}
\put(1213,749){\special{em:lineto}}
\put(1225,750){\special{em:lineto}}
\put(1237,750){\special{em:lineto}}
\put(1248,751){\special{em:lineto}}
\put(1260,752){\special{em:lineto}}
\put(1272,753){\special{em:lineto}}
\put(1284,754){\special{em:lineto}}
\put(1295,754){\special{em:lineto}}
\put(1307,755){\special{em:lineto}}
\put(1319,756){\special{em:lineto}}
\put(1331,756){\special{em:lineto}}
\put(1342,757){\special{em:lineto}}
\put(1354,757){\special{em:lineto}}
\put(1366,758){\special{em:lineto}}
\put(1377,759){\special{em:lineto}}
\put(1389,759){\special{em:lineto}}
\put(1401,760){\special{em:lineto}}
\put(1413,760){\special{em:lineto}}
\put(1424,761){\special{em:lineto}}
\put(1436,761){\special{em:lineto}}
\end{picture}

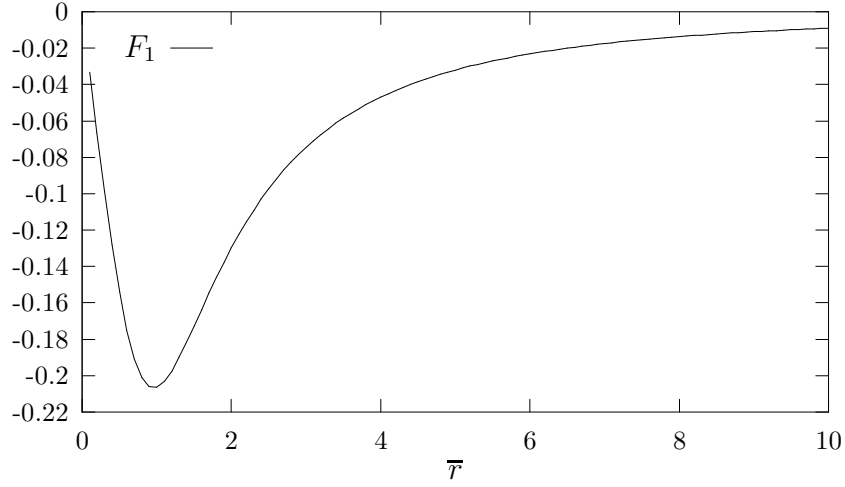
Fig2  The plot of right-hand ($W_{2}(M)$) and left-hand ($W_{1}(M)$) sides of 
Eq.(3) against $M$.
\vskip 1cm

The following conclusions can be made after considering the plots of the
above kind:

1. There are no equilibrium configurations whose the density is
larger than a certain value $\rho _{max} \sim 10^{16} gm/cm^{3}$ .

2. For  each value of $\rho < \rho _{max}$ there are two equilibrium
states (with $R > \alpha $  and $R < \alpha $). 

Are the configurations with largre masses stable?

The total energy of the degenerate gase is $E=E_{int} + E_{gr}$, where
$E_{int}$ is the intrinsic energy and  $E_{gr}$ is the gravitational
energy. The gravitational energy of a sphere is 
\begin{equation}
   E_{gr} = \int_{\infty}^{R} dM(r)\  \chi (r) \ M(r) ,
\end{equation} 
where
\begin{displaymath}
 \chi (r) = \int_{\infty}^r dr'\ (r')^{-2} (1 - 1/f) ,
\end{displaymath}
$\alpha  = 2GM(r)/c^{2}$, $f= (\alpha (r) ^3 +(r')^3)^{1/3}$,
\begin{displaymath}
M(r) =4 \pi  \int_{0}^{r} dr' \ \rho \ ( r')^{2} .
\end{displaymath}

The function $\chi (r)$ is approximately

\begin{equation}                                
   \chi (r) = (1/r)(1-\exp (-r/\alpha )).
\end{equation}

Therefore, at $p = const$  up to a constant of the order one
\begin{equation}
 E_{gr} = - \frac{G M^{2}}{R}(1 - \exp ( - R/\alpha ).
\end{equation}

The intrinsic energy $E_{int} =  \int u\  dM$, where $u$ is the energy per
the mass unit. For the used equation of state 
$u = K (\Gamma -1)^{-1} \rho ^{ \Gamma - 1}$ . Thus, up to a constants  of the
order of one
\begin{equation}
E=KM \rho ^{\Gamma -1} - G M^{5/3}\rho ^{1/3}[1-\exp(-QM^{-2/3}\rho ^{-1/3}],
\end{equation}
                                  
where $Q = c^2 /2G$. As an example, Fig.3 and Fig.4 show the plot of the 
function 
$E = E(\rho )$ for the nonrelativistic neutron gas of the mass $M=10^{36}\ gm$
and $M = 10^{33}\ gm$ (neutron stares) correspondingly.

\setlength{\unitlength}{0.240900pt}
\ifx\plotpoint\undefined\newsavebox{\plotpoint}\fi
\sbox{\plotpoint}{\rule[-0.200pt]{0.400pt}{0.400pt}}%
\special{em:linewidth 0.4pt}%
\begin{picture}(1500,900)(0,0)
\tenrm
\put(264,158){\special{em:moveto}}
\put(264,787){\special{em:lineto}}
\put(264,158){\special{em:moveto}}
\put(284,158){\special{em:lineto}}
\put(1436,158){\special{em:moveto}}
\put(1416,158){\special{em:lineto}}
\put(242,158){\makebox(0,0)[r]{-0.22}}
\put(264,215){\special{em:moveto}}
\put(284,215){\special{em:lineto}}
\put(1436,215){\special{em:moveto}}
\put(1416,215){\special{em:lineto}}
\put(242,215){\makebox(0,0)[r]{-0.2}}
\put(264,272){\special{em:moveto}}
\put(284,272){\special{em:lineto}}
\put(1436,272){\special{em:moveto}}
\put(1416,272){\special{em:lineto}}
\put(242,272){\makebox(0,0)[r]{-0.18}}
\put(264,330){\special{em:moveto}}
\put(284,330){\special{em:lineto}}
\put(1436,330){\special{em:moveto}}
\put(1416,330){\special{em:lineto}}
\put(242,330){\makebox(0,0)[r]{-0.16}}
\put(264,387){\special{em:moveto}}
\put(284,387){\special{em:lineto}}
\put(1436,387){\special{em:moveto}}
\put(1416,387){\special{em:lineto}}
\put(242,387){\makebox(0,0)[r]{-0.14}}
\put(264,444){\special{em:moveto}}
\put(284,444){\special{em:lineto}}
\put(1436,444){\special{em:moveto}}
\put(1416,444){\special{em:lineto}}
\put(242,444){\makebox(0,0)[r]{-0.12}}
\put(264,501){\special{em:moveto}}
\put(284,501){\special{em:lineto}}
\put(1436,501){\special{em:moveto}}
\put(1416,501){\special{em:lineto}}
\put(242,501){\makebox(0,0)[r]{-0.1}}
\put(264,558){\special{em:moveto}}
\put(284,558){\special{em:lineto}}
\put(1436,558){\special{em:moveto}}
\put(1416,558){\special{em:lineto}}
\put(242,558){\makebox(0,0)[r]{-0.08}}
\put(264,615){\special{em:moveto}}
\put(284,615){\special{em:lineto}}
\put(1436,615){\special{em:moveto}}
\put(1416,615){\special{em:lineto}}
\put(242,615){\makebox(0,0)[r]{-0.06}}
\put(264,673){\special{em:moveto}}
\put(284,673){\special{em:lineto}}
\put(1436,673){\special{em:moveto}}
\put(1416,673){\special{em:lineto}}
\put(242,673){\makebox(0,0)[r]{-0.04}}
\put(264,730){\special{em:moveto}}
\put(284,730){\special{em:lineto}}
\put(1436,730){\special{em:moveto}}
\put(1416,730){\special{em:lineto}}
\put(242,730){\makebox(0,0)[r]{-0.02}}
\put(264,787){\special{em:moveto}}
\put(284,787){\special{em:lineto}}
\put(1436,787){\special{em:moveto}}
\put(1416,787){\special{em:lineto}}
\put(242,787){\makebox(0,0)[r]{0}}
\put(264,158){\special{em:moveto}}
\put(264,178){\special{em:lineto}}
\put(264,787){\special{em:moveto}}
\put(264,767){\special{em:lineto}}
\put(264,113){\makebox(0,0){0}}
\put(498,158){\special{em:moveto}}
\put(498,178){\special{em:lineto}}
\put(498,787){\special{em:moveto}}
\put(498,767){\special{em:lineto}}
\put(498,113){\makebox(0,0){2}}
\put(733,158){\special{em:moveto}}
\put(733,178){\special{em:lineto}}
\put(733,787){\special{em:moveto}}
\put(733,767){\special{em:lineto}}
\put(733,113){\makebox(0,0){4}}
\put(967,158){\special{em:moveto}}
\put(967,178){\special{em:lineto}}
\put(967,787){\special{em:moveto}}
\put(967,767){\special{em:lineto}}
\put(967,113){\makebox(0,0){6}}
\put(1202,158){\special{em:moveto}}
\put(1202,178){\special{em:lineto}}
\put(1202,787){\special{em:moveto}}
\put(1202,767){\special{em:lineto}}
\put(1202,113){\makebox(0,0){8}}
\put(1436,158){\special{em:moveto}}
\put(1436,178){\special{em:lineto}}
\put(1436,787){\special{em:moveto}}
\put(1436,767){\special{em:lineto}}
\put(1436,113){\makebox(0,0){10}}
\put(264,158){\special{em:moveto}}
\put(1436,158){\special{em:lineto}}
\put(1436,787){\special{em:lineto}}
\put(264,787){\special{em:lineto}}
\put(264,158){\special{em:lineto}}
\put(850,68){\makebox(0,0){$\overline r$}}
\put(381,730){\makebox(0,0)[r]{$F_{1}$}}
\put(403,730){\special{em:moveto}}
\put(469,730){\special{em:lineto}}
\put(276,692){\special{em:moveto}}
\put(287,597){\special{em:lineto}}
\put(299,506){\special{em:lineto}}
\put(311,421){\special{em:lineto}}
\put(323,347){\special{em:lineto}}
\put(334,286){\special{em:lineto}}
\put(346,241){\special{em:lineto}}
\put(358,212){\special{em:lineto}}
\put(369,198){\special{em:lineto}}
\put(381,197){\special{em:lineto}}
\put(393,206){\special{em:lineto}}
\put(405,222){\special{em:lineto}}
\put(416,244){\special{em:lineto}}
\put(428,268){\special{em:lineto}}
\put(440,293){\special{em:lineto}}
\put(452,319){\special{em:lineto}}
\put(463,345){\special{em:lineto}}
\put(475,370){\special{em:lineto}}
\put(487,393){\special{em:lineto}}
\put(498,416){\special{em:lineto}}
\put(510,437){\special{em:lineto}}
\put(522,457){\special{em:lineto}}
\put(534,475){\special{em:lineto}}
\put(545,493){\special{em:lineto}}
\put(557,509){\special{em:lineto}}
\put(569,524){\special{em:lineto}}
\put(580,538){\special{em:lineto}}
\put(592,551){\special{em:lineto}}
\put(604,563){\special{em:lineto}}
\put(616,574){\special{em:lineto}}
\put(627,584){\special{em:lineto}}
\put(639,594){\special{em:lineto}}
\put(651,603){\special{em:lineto}}
\put(662,612){\special{em:lineto}}
\put(674,620){\special{em:lineto}}
\put(686,627){\special{em:lineto}}
\put(698,634){\special{em:lineto}}
\put(709,641){\special{em:lineto}}
\put(721,647){\special{em:lineto}}
\put(733,653){\special{em:lineto}}
\put(745,658){\special{em:lineto}}
\put(756,663){\special{em:lineto}}
\put(768,668){\special{em:lineto}}
\put(780,673){\special{em:lineto}}
\put(791,677){\special{em:lineto}}
\put(803,681){\special{em:lineto}}
\put(815,685){\special{em:lineto}}
\put(827,689){\special{em:lineto}}
\put(838,692){\special{em:lineto}}
\put(850,695){\special{em:lineto}}
\put(862,699){\special{em:lineto}}
\put(873,702){\special{em:lineto}}
\put(885,704){\special{em:lineto}}
\put(897,707){\special{em:lineto}}
\put(909,710){\special{em:lineto}}
\put(920,712){\special{em:lineto}}
\put(932,714){\special{em:lineto}}
\put(944,717){\special{em:lineto}}
\put(955,719){\special{em:lineto}}
\put(967,721){\special{em:lineto}}
\put(979,723){\special{em:lineto}}
\put(991,725){\special{em:lineto}}
\put(1002,726){\special{em:lineto}}
\put(1014,728){\special{em:lineto}}
\put(1026,730){\special{em:lineto}}
\put(1038,731){\special{em:lineto}}
\put(1049,733){\special{em:lineto}}
\put(1061,734){\special{em:lineto}}
\put(1073,736){\special{em:lineto}}
\put(1084,737){\special{em:lineto}}
\put(1096,738){\special{em:lineto}}
\put(1108,740){\special{em:lineto}}
\put(1120,741){\special{em:lineto}}
\put(1131,742){\special{em:lineto}}
\put(1143,743){\special{em:lineto}}
\put(1155,744){\special{em:lineto}}
\put(1166,745){\special{em:lineto}}
\put(1178,746){\special{em:lineto}}
\put(1190,747){\special{em:lineto}}
\put(1202,748){\special{em:lineto}}
\put(1213,749){\special{em:lineto}}
\put(1225,750){\special{em:lineto}}
\put(1237,750){\special{em:lineto}}
\put(1248,751){\special{em:lineto}}
\put(1260,752){\special{em:lineto}}
\put(1272,753){\special{em:lineto}}
\put(1284,754){\special{em:lineto}}
\put(1295,754){\special{em:lineto}}
\put(1307,755){\special{em:lineto}}
\put(1319,756){\special{em:lineto}}
\put(1331,756){\special{em:lineto}}
\put(1342,757){\special{em:lineto}}
\put(1354,757){\special{em:lineto}}
\put(1366,758){\special{em:lineto}}
\put(1377,759){\special{em:lineto}}
\put(1389,759){\special{em:lineto}}
\put(1401,760){\special{em:lineto}}
\put(1413,760){\special{em:lineto}}
\put(1424,761){\special{em:lineto}}
\put(1436,761){\special{em:lineto}}
\end{picture}

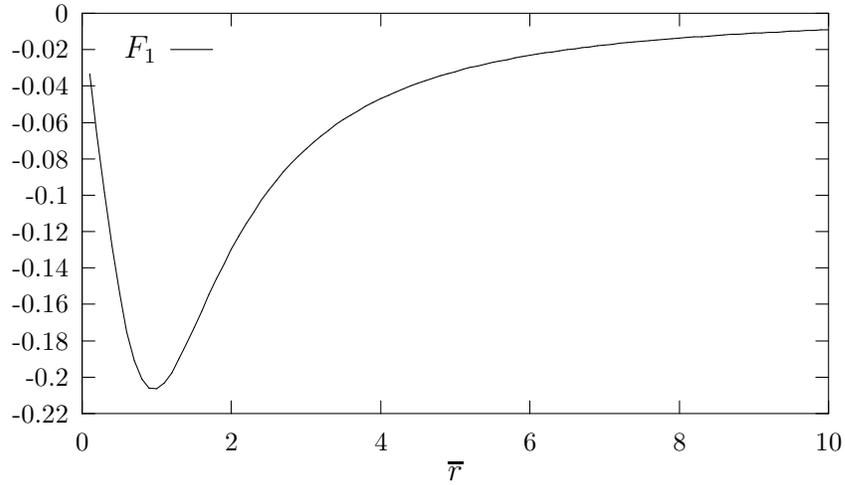
Fig. 3  The plot of the function $E = E(\rho)$ for the neutron configuration
of the mass $M = 10^{36}\ gm$.
\vskip 1cm 

\setlength{\unitlength}{0.240900pt}
\ifx\plotpoint\undefined\newsavebox{\plotpoint}\fi
\sbox{\plotpoint}{\rule[-0.200pt]{0.400pt}{0.400pt}}%
\special{em:linewidth 0.4pt}%
\begin{picture}(1500,900)(0,0)
\tenrm
\put(264,158){\special{em:moveto}}
\put(264,787){\special{em:lineto}}
\put(264,158){\special{em:moveto}}
\put(284,158){\special{em:lineto}}
\put(1436,158){\special{em:moveto}}
\put(1416,158){\special{em:lineto}}
\put(242,158){\makebox(0,0)[r]{-0.22}}
\put(264,215){\special{em:moveto}}
\put(284,215){\special{em:lineto}}
\put(1436,215){\special{em:moveto}}
\put(1416,215){\special{em:lineto}}
\put(242,215){\makebox(0,0)[r]{-0.2}}
\put(264,272){\special{em:moveto}}
\put(284,272){\special{em:lineto}}
\put(1436,272){\special{em:moveto}}
\put(1416,272){\special{em:lineto}}
\put(242,272){\makebox(0,0)[r]{-0.18}}
\put(264,330){\special{em:moveto}}
\put(284,330){\special{em:lineto}}
\put(1436,330){\special{em:moveto}}
\put(1416,330){\special{em:lineto}}
\put(242,330){\makebox(0,0)[r]{-0.16}}
\put(264,387){\special{em:moveto}}
\put(284,387){\special{em:lineto}}
\put(1436,387){\special{em:moveto}}
\put(1416,387){\special{em:lineto}}
\put(242,387){\makebox(0,0)[r]{-0.14}}
\put(264,444){\special{em:moveto}}
\put(284,444){\special{em:lineto}}
\put(1436,444){\special{em:moveto}}
\put(1416,444){\special{em:lineto}}
\put(242,444){\makebox(0,0)[r]{-0.12}}
\put(264,501){\special{em:moveto}}
\put(284,501){\special{em:lineto}}
\put(1436,501){\special{em:moveto}}
\put(1416,501){\special{em:lineto}}
\put(242,501){\makebox(0,0)[r]{-0.1}}
\put(264,558){\special{em:moveto}}
\put(284,558){\special{em:lineto}}
\put(1436,558){\special{em:moveto}}
\put(1416,558){\special{em:lineto}}
\put(242,558){\makebox(0,0)[r]{-0.08}}
\put(264,615){\special{em:moveto}}
\put(284,615){\special{em:lineto}}
\put(1436,615){\special{em:moveto}}
\put(1416,615){\special{em:lineto}}
\put(242,615){\makebox(0,0)[r]{-0.06}}
\put(264,673){\special{em:moveto}}
\put(284,673){\special{em:lineto}}
\put(1436,673){\special{em:moveto}}
\put(1416,673){\special{em:lineto}}
\put(242,673){\makebox(0,0)[r]{-0.04}}
\put(264,730){\special{em:moveto}}
\put(284,730){\special{em:lineto}}
\put(1436,730){\special{em:moveto}}
\put(1416,730){\special{em:lineto}}
\put(242,730){\makebox(0,0)[r]{-0.02}}
\put(264,787){\special{em:moveto}}
\put(284,787){\special{em:lineto}}
\put(1436,787){\special{em:moveto}}
\put(1416,787){\special{em:lineto}}
\put(242,787){\makebox(0,0)[r]{0}}
\put(264,158){\special{em:moveto}}
\put(264,178){\special{em:lineto}}
\put(264,787){\special{em:moveto}}
\put(264,767){\special{em:lineto}}
\put(264,113){\makebox(0,0){0}}
\put(498,158){\special{em:moveto}}
\put(498,178){\special{em:lineto}}
\put(498,787){\special{em:moveto}}
\put(498,767){\special{em:lineto}}
\put(498,113){\makebox(0,0){2}}
\put(733,158){\special{em:moveto}}
\put(733,178){\special{em:lineto}}
\put(733,787){\special{em:moveto}}
\put(733,767){\special{em:lineto}}
\put(733,113){\makebox(0,0){4}}
\put(967,158){\special{em:moveto}}
\put(967,178){\special{em:lineto}}
\put(967,787){\special{em:moveto}}
\put(967,767){\special{em:lineto}}
\put(967,113){\makebox(0,0){6}}
\put(1202,158){\special{em:moveto}}
\put(1202,178){\special{em:lineto}}
\put(1202,787){\special{em:moveto}}
\put(1202,767){\special{em:lineto}}
\put(1202,113){\makebox(0,0){8}}
\put(1436,158){\special{em:moveto}}
\put(1436,178){\special{em:lineto}}
\put(1436,787){\special{em:moveto}}
\put(1436,767){\special{em:lineto}}
\put(1436,113){\makebox(0,0){10}}
\put(264,158){\special{em:moveto}}
\put(1436,158){\special{em:lineto}}
\put(1436,787){\special{em:lineto}}
\put(264,787){\special{em:lineto}}
\put(264,158){\special{em:lineto}}
\put(850,68){\makebox(0,0){$\overline r$}}
\put(381,730){\makebox(0,0)[r]{$F_{1}$}}
\put(403,730){\special{em:moveto}}
\put(469,730){\special{em:lineto}}
\put(276,692){\special{em:moveto}}
\put(287,597){\special{em:lineto}}
\put(299,506){\special{em:lineto}}
\put(311,421){\special{em:lineto}}
\put(323,347){\special{em:lineto}}
\put(334,286){\special{em:lineto}}
\put(346,241){\special{em:lineto}}
\put(358,212){\special{em:lineto}}
\put(369,198){\special{em:lineto}}
\put(381,197){\special{em:lineto}}
\put(393,206){\special{em:lineto}}
\put(405,222){\special{em:lineto}}
\put(416,244){\special{em:lineto}}
\put(428,268){\special{em:lineto}}
\put(440,293){\special{em:lineto}}
\put(452,319){\special{em:lineto}}
\put(463,345){\special{em:lineto}}
\put(475,370){\special{em:lineto}}
\put(487,393){\special{em:lineto}}
\put(498,416){\special{em:lineto}}
\put(510,437){\special{em:lineto}}
\put(522,457){\special{em:lineto}}
\put(534,475){\special{em:lineto}}
\put(545,493){\special{em:lineto}}
\put(557,509){\special{em:lineto}}
\put(569,524){\special{em:lineto}}
\put(580,538){\special{em:lineto}}
\put(592,551){\special{em:lineto}}
\put(604,563){\special{em:lineto}}
\put(616,574){\special{em:lineto}}
\put(627,584){\special{em:lineto}}
\put(639,594){\special{em:lineto}}
\put(651,603){\special{em:lineto}}
\put(662,612){\special{em:lineto}}
\put(674,620){\special{em:lineto}}
\put(686,627){\special{em:lineto}}
\put(698,634){\special{em:lineto}}
\put(709,641){\special{em:lineto}}
\put(721,647){\special{em:lineto}}
\put(733,653){\special{em:lineto}}
\put(745,658){\special{em:lineto}}
\put(756,663){\special{em:lineto}}
\put(768,668){\special{em:lineto}}
\put(780,673){\special{em:lineto}}
\put(791,677){\special{em:lineto}}
\put(803,681){\special{em:lineto}}
\put(815,685){\special{em:lineto}}
\put(827,689){\special{em:lineto}}
\put(838,692){\special{em:lineto}}
\put(850,695){\special{em:lineto}}
\put(862,699){\special{em:lineto}}
\put(873,702){\special{em:lineto}}
\put(885,704){\special{em:lineto}}
\put(897,707){\special{em:lineto}}
\put(909,710){\special{em:lineto}}
\put(920,712){\special{em:lineto}}
\put(932,714){\special{em:lineto}}
\put(944,717){\special{em:lineto}}
\put(955,719){\special{em:lineto}}
\put(967,721){\special{em:lineto}}
\put(979,723){\special{em:lineto}}
\put(991,725){\special{em:lineto}}
\put(1002,726){\special{em:lineto}}
\put(1014,728){\special{em:lineto}}
\put(1026,730){\special{em:lineto}}
\put(1038,731){\special{em:lineto}}
\put(1049,733){\special{em:lineto}}
\put(1061,734){\special{em:lineto}}
\put(1073,736){\special{em:lineto}}
\put(1084,737){\special{em:lineto}}
\put(1096,738){\special{em:lineto}}
\put(1108,740){\special{em:lineto}}
\put(1120,741){\special{em:lineto}}
\put(1131,742){\special{em:lineto}}
\put(1143,743){\special{em:lineto}}
\put(1155,744){\special{em:lineto}}
\put(1166,745){\special{em:lineto}}
\put(1178,746){\special{em:lineto}}
\put(1190,747){\special{em:lineto}}
\put(1202,748){\special{em:lineto}}
\put(1213,749){\special{em:lineto}}
\put(1225,750){\special{em:lineto}}
\put(1237,750){\special{em:lineto}}
\put(1248,751){\special{em:lineto}}
\put(1260,752){\special{em:lineto}}
\put(1272,753){\special{em:lineto}}
\put(1284,754){\special{em:lineto}}
\put(1295,754){\special{em:lineto}}
\put(1307,755){\special{em:lineto}}
\put(1319,756){\special{em:lineto}}
\put(1331,756){\special{em:lineto}}
\put(1342,757){\special{em:lineto}}
\put(1354,757){\special{em:lineto}}
\put(1366,758){\special{em:lineto}}
\put(1377,759){\special{em:lineto}}
\put(1389,759){\special{em:lineto}}
\put(1401,760){\special{em:lineto}}
\put(1413,760){\special{em:lineto}}
\put(1424,761){\special{em:lineto}}
\put(1436,761){\special{em:lineto}}
\end{picture}

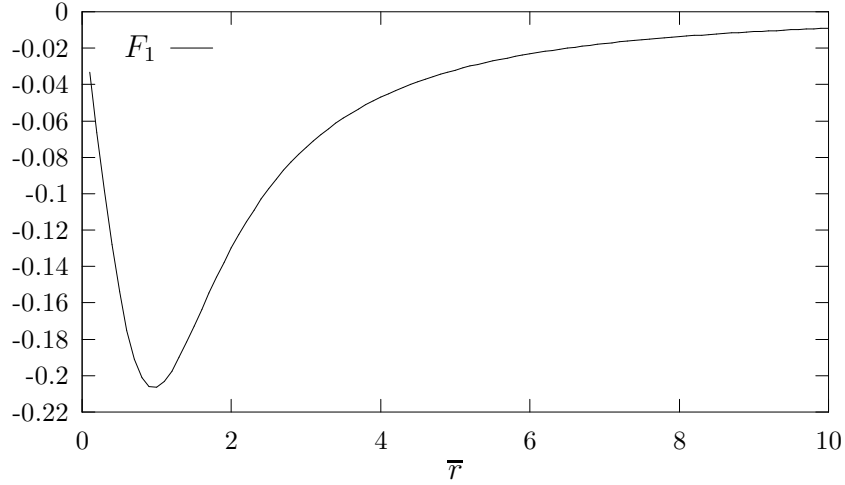
Fig.4  The plot of the function $E = E(\rho)$ for the neutron star of the 
mass $M = 10^{33} \ gm$.
\vskip 1 cm

The analysis of  such plots show that the function $E = E (\rho)$ has the 
minimum. Thus, the above equilibrium states of large masses  are stable.

  The gravitational potential on the surface of a stable massive 
configuration of the degenerate fermion gas of the order of
\begin{equation}
   V = (G M /R) [1 - \exp (-R/\alpha )]
\end{equation}
It follows from the virial theorem that the above objects of large masses 
(their $R \ll \alpha $) are the ones with low temperatures . They,
probably, refer to "dark" matter of the Universe. If their luminosity are
caused by an accretion, then the Eddington limit of luminosity is
approximately
\begin{equation}
 { \cal L} = {\cal L}^{0} _{Edd} [ 1- \exp (-R / \alpha )],
\end{equation}
where ${\cal L}^ {0}_{Edd} = 1 \cdot 10^{39} M \ erg/s $. Hence, if 
$R/\alpha \ll 1 $, their luminosity  ${\cal L} \ll {\cal L}^{0}_{Edd} $.


\begin{thebibliography}{10}
\bibitem{ver}      L.V.Verzub,  Phys.Lett. A , v.156 , 404 (1991)
\bibitem{teuk}     S.L.Shapiro and Teukolsky , Black Holes, White Drafts and
Neutron Stars, (1983)
\bibitem{chan} S.Chandrasekhar , Mon. Not. Roy. Astron. Soc. v.95 , 226 (1935)
\end{thebibliography}
\end{document}